# Radiation damage of polyethylene exposed in the stratosphere at an altitude of 40 km

Alexey Kondyurin, Irina Kondyurina, Marcela Bilek

Applied and Plasma Physics, School of Physics A28, University of Sydney, Sydney, NSW 2006, Australia

**Abstract**

Low Density Polyethylene (LDPE) films were exposed at an altitude of 40 km over a 3 day NASA stratospheric balloon mission from Alice Springs, Australia. The radiation damage, oxidation and nitration in the LDPE films exposed in stratosphere were measured using ESR, FTIR and XPS spectroscopy. The results were compared with those from samples stored on the ground and exposed in a laboratory plasma. The types of free radicals, unsaturated hydrocarbon groups, oxygen-containing and nitrogen-containing groups in LDPE film exposed in the stratosphere and at the Earth's surface are different. The radiation damage in films exposed in the stratosphere are observed in the entire film due to the penetration of high energy cosmic rays through their thickness, while the radiation damage in films exposed on the ground is caused by sunlight penetrating into only a thin surface layer. A similarly thin layer of the film is damaged by exposure to plasma due to the low energy of the plasma particles. The intensity of oxidation and nitration of LDPE films reflects the difference of atmospheric pressure on the ground and in the stratosphere. The high-density radiation damage of the LDPE films above the ozone layer in the stratosphere is caused by primary cosmic rays as well as collision induced cosmic ray air showers, and is consistent with the measured flux of cosmic radiation. The results show, that stratospheric flights can be used to simulate the effects of space environments during interplanet space flights for the purposes of investigating the degradation of polymer materials.

## Introduction

The environment in free space is destructive for materials. In space conditions such as high vacuum, extreme temperature cycling and exposure to high energy cosmic rays including protons and $\alpha$-particles, electrons, $\gamma$-rays, X-rays, UV and VUV light, damage the materials from which space ships, space stations and satellites are constructed. A number of investigations of exposure of different materials are being carried out on the International Space Station and other satellites to acquire knowledge of the destructive processes in space. These experiments are quite expensive and samples are limited in mass and volume by the cost of transporting materials into space and by the volume of the space station, while the exposure time is limited by the duration of the space flight.

Another way of investigating damage to materials in space is to expose the samples during stratospheric balloon flights. When a balloon is above the ozone layer, the irradiations to which the materials are exposed are essentially those encountered in free space. The intensity of space irradiations is lower, than for satellites in Low Earth Orbit (LEO) due to absorbance in top atmosphere, but higher than on the Earth's surface. The most important factor for material destruction in LEO is the flux of atomic oxygen, which is absent in the stratosphere as well as in deep space. This gives us the opportunity of using stratospheric observation to assess the potential damage to materials in deep space. On the other hand, the deep space environment contains no oxygen while the stratospheric conditions include a low pressure of oxygen. This combination of space factors in stratospheric flight gives us the freedom to separately consider the damaging processes in polymer materials caused by different factors. An additional advantage of the stratospheric flight experiments is low cost, which is very important for modern space programs.

Exposure in space is expected to strongly impact on polymer materials due to the combined effects of low atmospheric pressure, cosmic rays, high intensity UV radiation including short wavelength UV, diurnal temperature variations and other effects associated with solar irradiation on the chemical processes in polymeric materials. Since this combination of conditions cannot be adequately simulated in the laboratory, it is difficult to predict the impact. Such knowledge is particularly important in the development of polymers for long term structural applications in deep space, such as the polyethylene sheets proposed to shield high energy cosmic radiation during long term space flights such as the Mars mission [1, 2].

Polyethylene film impregnated with Zylon fibers is used for the shell of the stratospheric balloon [3, 4]. The mechanical properties of the polyethylene shell satisfy the requirements of balloon deployment and stability in stratosphere [5-8]. For long duration flights the polyethylene film will be exposed under stratospheric conditions for some months. Long term flights require high structural strength combined with minimal mass. The optimisation of strength/mass requires prediction of the destruction processes in a polyethylene shell for a range of the stratospheric conditions.

The effect of the LEO environment on the polyethylene has been investigated in space and in a simulated free space environment [9-11]. The main effect is an erosion in space of the polyethylene film due to a flux of atomic oxygen with ~5 eV kinetic energy. The rate of etching for High Density Polyethylene (HDPE) is 3.5-3.7 x $10^{24}$ $cm^3$/atom. A second effect is oxidation of the surface layer by atomic oxygen. Other effects of high energy space irradiation on polyethylene were not detectable due to their relatively low intensity.

Destruction of different polymers in free space has been intensively studied, in experiments outside a space ship on LEO and in laboratory experiments [for example 12-19]. These studies showed, that destruction of epoxy composites, HDPE, LDPE, Kapton, Teflon and FEP Teflon, Tedlar, Mylar, fluorinated polyimides, polyurethanes, polysulfone and others during flight on low Earth orbit is primarily due to atomic oxygen bombardment [20]. The results are loss of mass, dehydration processes, the formation of an amorphous carbon layer, the formation of cracks and craters on the surface, and a decrease in the durability of the polymer.

The influence of individual space factors on polymers has been studied in more detail in laboratory experiments. The atomic oxygen, electron and ion fluxes, and the UV light intensity depend on the sun's activity and space ship mission conditions. The NASA database [21] can be consulted to identify space orbit conditions for laboratory simulation. Usually, plasma discharges, ion and atomic beams, and UV light from Krypton, Xenon, Mercury and Hydrogen lamps are used for simulating the space environment in a laboratory. However, no single method provides complete agreement on the effect of space irradiation in simulated and real space environments.

Polymer degradation after exposure in real and simulated free space environment has been observed by optical, electron and atomic force microscopes as an increase in surface roughness; by XPS and FTIR spectroscopy as oxidation of surface layer; by mechanical and thermo-mechanical methods as a decrease in the strength and an increase in the Young's modulus and the glass transition temperature [22-30]. There does not appear to be, however, any detailed investigation of the effects of stratospheric conditions on polymer materials. This paper reports an investigation of the effect of the stratospheric environment on polymer materials compared with that in a laboratory simulated environment. This will facilitate better prediction of polymer behaviour for future stratospheric missions and give a better understanding of the polymer degradation processes under free space conditions.

## Experiment

*Stratospheric flight*

This flight experiment was a part of the NASA scientific balloon flight program realised at the NASA stratospheric balloon station in Alice Springs, Australia. A flight cassette with the polymer samples was installed on the` 1200 kg payload carrying the telescope of the Tracking and Imaging Gamma-Ray Experiment (TIGRE). The payload was lifted with a "zero-pressure" helium filled stratospheric balloon of diameter 300 m. The Columbia Scientific Balloon Facility (CSBF) provided the successful launch, stable flight telemetry and gentle landing of the payload.

Three cassettes with polymer samples were prepared (Fig.1). The cassette consists of an aluminium base covered with paint filled with ZnO particulars (Dulux, Australia). The first datalogger with temperature sensor and microprocessor data storage unit (EL-USB-1, model 23039-50, USA) was placed in an aluminium cylinder, sealed with aluminium disks and glued hermetically with Araldite epoxy resin. The second datalogger was placed under the base in the black sleeve. The complete description of the cassette is presented in [31]. The top of the cassette was covered with two layers of LDPE films of 0.05 mm thickness and 10x10 $cm^2$ area purchased from Goodfellow (UK). The films were stretched and fixed with 6 stainless steel screws. The gap between films was not regulated, but some space between films and between second film and base was clearly observed. A ground control cassette was prepared at the same time with the same samples as the flight cassette. A third cassette was prepared and kept in a refrigerator at +2-3$^0$C all time.

At 1 am on the 16 April, the flight and ground control cassettes were taken from the refrigerator. This time and all further references to time in the text refer to the local, Northern Territory, time. The flight cassette was fixed on the payload to the GPS antenna bar and moved out to the airstrip. The balloon was launched from Alice Springs Seven Mile airport, Northern Territory, Australia, on the 16 April, 2010, at 9:00 am. After 2 hours the balloon had risen to 40 km altitude.

Over the next three days the altitude of the balloon varied between 40 km (day time) and 35 km (night time). The geographical coordinates of the balloon, it's altitude, the pressure, the temperature and the signal from the video camera were monitored with telemetry. The payload was rotated by a motor at a rate of 1 turn per 4 minutes during two days of flight. On the 3rd day of flight, the rotation was stopped and the orientation of the payload was no longer controlled. After 3 days of flight, the payload was separated from the balloon and descended by means of a parachute. After 3 hours of descent the payload landed about 100 km to the west from Longreach, Queensland, Australia (latitude 240 2.71' S, longitude 1430 54.5' E), 990 km from the launch site, at 18:11 on the 18 April, 2010. The speed at landing was about 4-5 m/sec. At landing, the payload fell on the side where the cassette was located with the top of the instrumentation cylinder with part of LDPE film in contact with the dry soil. The part of film which had been contaminated by soil was excluded from analysis. The day after landing the payload was found and transported to Longreach airport where the flight cassette was removed and stored in a refrigerator (+2$^0$C).

The ground control cassette was placed outside the NASA base shed at the time of launch and exposed to sun light during the balloon flight. The ground control cassette was placed in the refrigerator at the same time as the flight cassette was removed from the payload and placed in the refrigerator. The two cassettes were refrigerated for transport to Sydney in and prior to analysis. Data loggers recorded the temperature on both cassettes during the waiting period before launch, the flight and during transportation.

Pressure, temperature and altitude were recorded and logged during the flight. These data were sent to the ground station using telemetry. The air temperature, plotted in Figure 1, was measured by a sensor placed in a white painted box at the bottom of the payload. It showed that the temperature decreased after launch from +20$^0$C on the ground reaching -76.7$^0$C at an altitude of 17600 m. During the flight at 40 km altitude, the temperature remained in the range -20 to +5$^0$C during the day and -30 to - 45$^0$C at night.

The daytime temperature varied periodically with time. The oscillations with period of 4 mins corresponded to the rotation of the payload, the high temperature corresponding to an orientation of the temperature sensor to the sun, where sunlight heated the sensor. The low temperature corresponds to the temperature of the air. On the 3$^{rd}$ day, the rotation of the payload stopped and the temperature varied between -22 and +13$^0$C randomly depending on the orientation of the payload with respect to the sun.

The temperature was also measured inside the cylinder above the base plate and in the sleeve below the aluminium base plate (Fig.1.) The temperature increased to 26$^0$C in the cylinder and to 20.5$^0$C in the sleeve due to solar irradiation at higher altitude. At night time, the temperature decreased below -40$^0$C. Temperatures lower than -40$^0$C were not recorded as this is the lower limit of the measurement range for our thermometers.

At 7:25 am, on the second day, thermometer 2 in the sleeve stopped recording. The maximum temperature of the flight cassette (+32.5$^0$C) was observed on the third day. The low (-38$^0$C) temperature observed on the third day corresponds to a descent of the payload to an altitude of 20 km.

The temperature of the ground control cassette was consistently higher than the flight cassette. The maximum temperature of the ground cassette was +37.5$^0$C.

The pressure decreased with elevation of the balloon. At daytime, the balloon was heated by solar irradiation. The altitude of the balloon remained in the range 38- 39 km during the first and second days of flight and then increased to 40 km on the 3rd day of flight. The pressure at daytime decreased to 2.5-2.1 Torr while at night time, the balloon cooled, reducing the lift on the balloon, and the altitude decreased to the range 34-35 km with a consequent increase in pressure to 5 Torr.

During the flight (16-18 of April, 2010) the space weather was monitored using Space Weather Prediction Center at National Oceanic and Atmospheric Administration (NOAA/SWPC) Boulder, CO USA data received from the GOES-13 satellite on geostationary orbit. The electron flux and proton flux correspond to low level solar activity (Table 1). The X-ray intensity was $10^{-8}$-$10^{-9}$ W/m$^2$ in the 0.5-4 Å wavelength range, (2.0-0.2)x$10^{-7}$ W/m$^2$ for the 1-8 Å wavelength range. The U.S. Dept. of Commerce, NOAA, Space Weather Prediction Center registered no space weather events during the flight days.

Table 1. Fluxes of electrons and protons on flight days of 16-18, April, 2010 measured at the GEOS-13 spacecraft on Geostationary orbit (NOAA data).

| Particles | Energy MeV | Average flux in a day Particles/cm$^2$/day/sr | | | Sum of particles for 3 days on the surface |
|---|---|---|---|---|---|
| | | Day 1 | Day 2 | Day 3 | Particles/cm$^2$ |
| Electrons | > 0.8 | 8.83E+08 | 7.86E+08 | 7.89E+08 | 7.72E+09 |
| | > 2 | 1.35E+07 | 1.57E+07 | 1.80E+07 | 1.48E+08 |
| Protons | > 1 | 2.25E+05 | 2.39E+05 | 2.76E+05 | 2.32E+06 |
| | > 10 | 1.74E+04 | 1.72E+04 | 1.78E+04 | 1.65E+05 |
| | > 100 | 7.07E+03 | 6.72E+03 | 7.16E+03 | 6.58E+04 |

The solar irradiance data were taken at the Laboratory for Atmospheric and Space Physics (LASP). The data was received from the SORCE spacecraft in Earth orbit. This dataset included daily averaged Solar Spectrum Irradiance (SSI) measurements from three instruments: Spectral Irradiance Monitor (SIM) (which gives data for the 310-2400 nm wavelengtgh range); SOLar STellar Irradiance Comparison Experiment (SOLSTICE) (115-310 nm), XUV Photometer System (XPS) (0.1-40 nm). The spectra during all days of the stratospheric flight of 16, 17 and 18 of April 2010 are similar. The total level of the solar irradiation was 1361.09±0.01 w/m$^2$ over all days of the flight.

During the flight the cosmic ray intensity was measured using a Compton telescope: the Tracking and Imaging Gamma Ray Experiment (TIGRE) with multi-layers of thin silicon strip detectors to convert and track gamma-ray events [32].

The level of radiation on the ground is mostly provided by muons and corresponds to about 1000 counts per 1.31 sec in a hydrocarbon composition scintillator of 81.3x81.3x0.635 $cm^3$ size and 1 $g/cm^3$ density. The cosmic ray intensity in the stratosphere is about 5000 counts per 1.31 sec for the same scintillator. This is due to high energy (>1 MeV) cosmic ray protons, electrons, X-rays and γ-rays. These irradiations and UV/VUV light penetrated into the samples during the balloon flight at 40 km altitude, which is above ozone layer (20-25 km).

*Plasma treatment of LDPE film*

A helicon wave radio-frequency (13.56 MHz) plasma was used for treatment of LDPE films. The plasma power was 100 W with a reflected power of 12 W when matched. The base pressure of the vacuum chamber was $10^{-5}$ Torr and the pressure of nitrogen during the implantation process was $2 \cdot 10^{-3}$ Torr. The samples were placed on a floating metal substrate holder. The plasma parameters were measured using a Langmuir probe consisting of a 0.20 mm diameter tungsten wire passing through a sintered alumina ceramic tube inserted into a stainless steel tube which was grounded to the chamber. The probe was positioned above the substrate holder. Hiden Analytical Ltd. electronics and software were used for measurement and analysis of the data. The floating potential above the substrate electrode was -16 V during plasma treatment and the plasma density was $1.2 \times 10^9$ ions/$cm^3$.

*Analysis of LDPE film*

Electron spin resonance spectra (ESR) were recorded on a Bruker Elexsys E500 EPR spectrometer operating in X band with a microwave frequency of 9.75 GHz and a central magnetic field of 3480 G, at room temperature. The spectrometer was calibrated using a weak pitch sample in KCl and also with DPPH (α,α'–diphenyl-β-picrylhydrazyl). The LDPE films were rolled and placed into a glass sample tube to record the spectra. The spectrum of the tube without the LDPE film was recorded before the measurement.

Fourier Transform Infrared (FTIR) Attenuated Total Reflectance (ATR) spectra from the samples were recorded using a Digilab FTS7000 FTIR spectrometer fitted with an ATR accessory (Harrick, USA) with trapezium germanium crystal and an incidence angle of 45°. To obtain sufficient signal/noise ratio and spectral resolution, we used 500 scans and a resolution of 4 $cm^{-1}$. This technique gives spectra for a thin (400-800 nm) surface layer of the film. The spectra from both surfaces of the film were recorded.

XPS measurements were made with a Specs spectrometer (Specs, Germany), equipped with an Al X-ray source operating at 200 W and an electrostatic monochromator, a hemispherical analyser and a 9 channel line delay detector. Survey spectra were acquired for binding energies in the range 50 to 1200 eV, using a pass energy of 30 eV and spectra of the C 1s, O 1s and N 1s region were then acquired at a pass energy of 23 eV with 10 scans to obtain higher spectral resolution and reduce the noise level.

## Results

LDPE films exposed in stratosphere are visually similar to the films exposed on Earth as well as to the virgin films. There was no visible damage and the colour of the films is not changed.

Figure 2 shows that the ESR spectra from the films exposed in the stratosphere, on Earth and from the virgin film were different. The virgin film has free radicals with a g-factor of 2.0025 caused by environmental radiation. The film exposed in the ground cassette has an additional signal with g-factor of 2.011 due to solar radiation damage. The film exposed in the stratosphere has a strong signal with a g-factor of 2.0025 and signals with g-factors of 2.0061, 2.011, 2.017 and 2.021. The differences in g-factor indicate that different types of free radicals are created by stratospheric exposure [33].

The ESR signals are weak corresponding to a low concentration of the free radicals in the films. A low concentration can be expected because the films were stored at room temperature and the most of the free radicals resulting from radiation damage recombine to give more stable chemical groups. The exposure time (3 days) was not long enough to significantly damage the macromolecules in comparison with the usual lifetime of such films in sunlight (1-3 years) [34].

FTIR ATR spectra show radiation damage and oxidation of the films on the ground, in the stratosphere and after plasma treatment (Fig.3). The spectra of the exposed films show lines attributed to oxygen containing group ν(C=O) vibrations and the vibrations of unsaturated hydrocarbon groups γ(=CH), such as the vinyl (910 $cm^{-1}$), vinylidene (888 $cm^{-1}$) and vinylene (965 $cm^{-1}$) groups. The vinylidene group vibration line is only observed in the spectra of initial unexposed LDPE film. This group results from the branched polymerisation reactions in the polymerisation of LDPE. The line intensity of the vinylidene groups does not increase with exposure of the films on the ground or in the stratosphere, while the vinylene group absorbance increases slightly. The intensity of this line increases significantly in the spectra of the film after plasma treatment. Most of the changes in the absorbance of unsaturated groups in the films exposed on the ground and in the stratosphere are observed in the vinyl group concentrations. The absorbance of the 910 $cm^{-1}$ line of this group was used for a quantitative analysis of the unsaturated group concentrations. The absorbance of the 910 $cm^{-1}$ line normalised on the 1462 $cm^{-1}$ line due to methylene groups is proportional to the relative concentration of vinyl groups (Fig.4). Assuming that the average molecular mass of LDPE is 100000, the relative absorbance of the 910 $cm^{-1}$ line in the spectra of virgin LDPE film corresponds to about 1 vinyl group per macromolecule.

The concentration of vinyl groups in the exposed surfaces depends on the location of the surfaces of the film. The concentrations of vinyl groups in both surfaces of the second film, and rear surface of first film exposed on the ground, are similar to that of the virgin unexposed film. A high concentration (4.2 per macromolecule) is observed only on the front surface of first film exposed on the ground, which was directly irradiated by sunlight. The highest concentration of vinyl groups (6.8 per macromolecule) is observed on the front surface of the film exposed in the stratosphere. However, the internal surfaces of the films exposed in the stratosphere also have a high concentration of vinyl groups (2.1 per macromolecule), despite the fact that the first LDPE film shielded these surfaces from direct sunlight.

The concentration of vinyl groups increases with treatment time in the plasma. However, the vinyl group line is observed only for the front surface of the first film. The rear surface of the first film and both surfaces of the second film have the same concentration of vinyl groups as the virgin film. After 30 min of plasma treatment the concentration of vinyl groups (4.2 per macromolecule) in the front surface becomes similar to the concentration of the vinyl groups in the front surface of the film exposed on the ground, but this is lower than the concentration of vinyl groups on the front surface of the film exposed in the stratosphere.

A number of lines attributed to vibrations of oxygen containing groups are observed in the spectra of films after stratospheric, ground and plasma exposure. Hydroxyl group stretch vibrations ν(O-H) and ν(N-H) are observed in the region 3500-3200 $cm^{-1}$ and the complex band of the ν(C=O) carbonyl group vibrations is observed in the region 1750-1600 $cm^{-1}$. This band is a combination of overlapping individual lines attributed to specific groups such as to esters at 1640-1650 $cm^{-1}$, to unsaturated ketones at 1685-1695 $cm^{-1}$, carboxyl acid at 1700-1715 $cm^{-1}$, saturated ketones at 1720-1730 $cm^{-1}$, aldehydes at 1740 $cm^{-1}$, and to lactones at 1750-1765 $cm^{-1}$ [35].

The shape of the carbonyl band is different for the LDPE film exposed in the stratosphere, on the ground and in the plasma. To identify individual peaks associated with different groups, the spectra were analysed using a deconvolution procedure and the second derivative. This procedure gives the number and position of the individual peaks, which facilitated fitting of experimental spectra by individual Gaussian functions. Examples of the fitting of the FTIR ATR spectra for the front surfaces of the films exposed in the stratosphere and on the ground are presented in Fig.5. The strong lines observed are interpreted as ν(C=O) vibrations in lactone (about 1756 $cm^{-1}$), aldehyde (about 1740 $cm^{-1}$), carboxyl acids (about 1715 $cm^{-1}$), unsaturated ketones (about 1675 $cm^{-1}$) and esters (1630 and 1650 $cm^{-1}$). The weak lines at 1607, 1593 and 1583 $cm^{-1}$ are attributed to stretch ν(C=C) vibrations of unsaturated groups. The results of fitting the ATR spectra, presented in Table 2, show that the relative intensities of oxygen containing groups are different in the films exposed in the stratosphere and on the ground. The line corresponding to the esters is stronger in the film exposed in the stratosphere while the aldehyde line is stronger in the film exposed on the ground. The spectra of plasma treated LDPE show strongly overlapping lines in this region which cannot be easily interpreted.

Table 2. Area (in arbitrary units) of carbonyl group lines in the FTIR ATR spectra of LDPE films exposed in the stratosphere and on the ground.

| Group | Ester | Ester | Unsaturated ketone | Carboxyl acid | Aldehyde | Lactone |
|---|---|---|---|---|---|---|
| Line, $cm^{-1}$ | 1630 | 1650 | 1675 | 1715 | 1740 | 1756 |
| On the ground | 3.9 | 16.2 | 24.6 | 23.8 | 17.4 | 14.0 |
| In the stratosphere | 2.8 | 32.1 | 28.1 | 26.8 | 6.5 | 3.7 |

The total concentrations of carbonyl groups in films exposed in the stratosphere and on the ground are different. The integrated intensity in the 1800-1600 $cm^{-1}$ region divided by the integrated intensity of 1462 $cm^{-1}$ δ(CH) bending vibrations of methylene group which is proportional to the relative total concentration of carbonyl groups, is shown in Fig.6. There is a low concentration of carbonyl groups in the virgin LDPE film. These groups in virgin film are a result of radiation damage and oxidation caused by environmental effects during storing under atmospheric conditions. The highest concentration of carbonyl groups is observed in the external surface of the first film exposed on the ground. The concentration of carbonyl groups in the internal surface of first film and in both surfaces of second film exposed on ground is the same as the concentration in the virgin film. The concentration of carbonyl groups in all the surfaces of the films exposed in the stratosphere is higher than in the virgin film. However, the total concentration of carbonyl groups in external surface of film exposed in the stratosphere is two times lower than in the film exposed on the ground.

The concentration of carbonyl groups in the plasma treated films increases with the duration of the plasma treatment, approaching an absorbance of 0.1 asymptotically. This concentration, after 30 min of plasma treatment is lower than that in the ground-exposed film (0.132) but higher than in the stratospheric-exposed film (0.066).

Strong lines of the ν(C-O) vibrational spectrum are also observed in a region of 1100-1000 $cm^{-}1$ for exposed films. These line can be attributed to ν(C-O) vibrations in ether, ester, carboxyl and hydroxyl groups. For a quantitative analysis the absorbance of the 1031 $cm^{-1}$ line was normalised to the absorbance of the 1462 $cm^{-1}$ line of the methylene group.

In all surfaces of the films exposed in the stratosphere, the absorbance of C-O groups at 1031 $cm^{-1}$ is higher than in the virgin film, the highest absorbance being observed in the external surface of first film exposed in the stratosphere (Fig.3). Lower absorption is observed in the external surface of the first film exposed on ground, the absorbance in internal surfaces of these films being the same as for the initial film.

A strong broad line at 3341 $cm^{-1}$, attributed to ν(O-H) vibrations in a hydroxyl group, is observed in the spectra of the films treated by the plasma, and a high intensity ν(O-H) line is observed in the spectrum of the external surface of the first film exposed on ground. (Fig.3). The intensity of ν(O-H) line in the spectra of the internal surfaces of the first and second films exposed on the ground is low. A low intensity ν(O-H) line is also observed in the spectrum of the external surface of first film exposed in stratosphere, while the spectra of other surfaces of films exposed in the stratosphere do not show this line.

XPS spectra of the LDPE films exposed in the stratosphere, on the ground and in the plasma show the presence of oxygen and nitrogen, while the spectrum of the virgin LDPE film shows essentially only carbon (Table 3). The 285 eV C1s line in the spectrum of the virgin film is single and corresponds to carbon atoms in carbon-carbon bonds. The C1s line in XPS spectra of exposed films has a complex shape. The second derivative of the spectra shows a presence of 4 individual peaks. These peak positions were used for a fitting Gaussians to the C1s lines in the spectra for films exposed on the ground, in the stratosphere and treated by plasma. The fitted peaks at 286.1, 287 and 289 eV in the spectra correspond to carbon atoms in C-O, C-N, C=O and C=N bonds (Fig.8). A C1s peak of similar shape is observed for Kapton F exposed in LEO [36].

The concentration of oxygen in the films exposed on the ground is about 2 times higher, than in the films exposed in the stratosphere (Table 3). The amount of oxygen in the plasma treated film is between that in films exposed in the stratosphere and on the ground. The O1s line of oxygen is a doublet and can be fitted with two individual Gaussian peaks at 531.9 eV and 533.2 eV. The peak at 531.9 eV is attributed to an oxygen atom in a double bond with a carbon atom and the peak at 533.2 eV is attributed to oxygen in a C-O bond. The relative concentration of oxygen as C=O in films exposed in the stratosphere and in the plasma is higher than for the film exposed on the ground (Table 4).

The film treated for 30 min in nitrogen plasma exhibits a strong N1s line at 400.5 eV, corresponding to nitrogen bonded to carbon (Fig.9). The nitrogen that penetrates into the film during plasma treatment bonds with carbon atoms giving stable nitrogen-containing groups. The films exposed in stratosphere and on the ground show a much lower nitrogen concentrations. The nitrogen concentration in these films, however, is higher than in the virgin film; and the intensity of the N1s line in these spectra is higher than the noise level. The nitrogen concentration in the film exposed on the ground is about 2 times higher than for that exposed in the stratosphere.

Table 3. Intensity of C1s, O1s and N1s peaks, as a percentage of the total area of the XPS peaks, for all the elements recorded in the XPS spectra of LDPE exposed in the stratosphere (flight), on the ground and in a nitrogen plasma for 30 min.

| XPS peak | Virgin | Flight | Ground | Plasma |
|---|---|---|---|---|
| C1s | 99.9% | 93.9% | 86.9% | 77.6% |
| N1s | 0% | 0.3% | 0.6% | 15.0% |
| O1s | 0.1% | 5.8% | 12.5% | 7.4% |

Table 4. Relative intensity of fitted O1s peaks at 533.2 and 531.9 eV in the XPS spectra of LDPE associated with oxygen atoms in C-O and C=O groups for films exposed in the stratosphere (flight), on the ground and in a nitrogen plasma for 30 min.

| O1s peak | Virgin | Flight | Ground | Plasma |
|---|---|---|---|---|
| C=O, 531.9 eV | 0% | 66% | 41% | 66% |
| C-O, 533.2 eV | 100% | 34% | 59% | 34% |

**Discussion**

The stratosphere is characterised by a low atmospheric pressure, high temperature variations, high energy particles, X-rays, gamma-rays, and UV light including short wavelength UV light. Some components of the primary irradiation from space are absorbed at 80-100 km altitude and lower, reducing the primary irradiation at the 40 km balloon altitude. On the other hand, secondary irradiation due to collisions of high energy particles with gases of the residual atmosphere are generated at this altitude.

Most of the space irradiation is absorbed in the dense atmosphere at 20-25 km altitudes so that the level of radiation in stratosphere at an altitude of 40 km is expected to be higher than that on the ground.

High energy irradiation destroys polyethylene macromolecules so that hydrogen or carbon atoms leave their positions and move far from the mother macromolecule. As a consequence of this displacement, all the bonds of the displaced atom in macromolecule are broken and a number of the valence electrons belonging to displaced atom and its neighbours become unpaired, one displaced hydrogen atom giving 2 free radicals while one displaced carbon atom gives 4 free radicals in the mother macromolecule and 4 free radicals at new positions.

The distribution of the displaced atoms depends on the kind and energy of the penetrating particle. Fig.10 shows the free radical profiles generated by incoming protons with kinetic energies of 1 and 10 MeV. Protons with energies of 1 MeV or less penetrate into first film and stop there to cause destruction in the first surface layer of the film. The rear surface of the first film remains undamaged. Protons with energies of 10 MeV and higher penetrate and cause damage throughout both films.

UV light, including the short wavelength component, is strongly absorbed by polyethylene so that only the first surface layer of polyethylene is affected by UV while the rear surface of the first film and both surfaces of the second film are not illuminated and cannot be affected. The X-rays and $\gamma$-rays as well as high energy electrons penetrate through the films and can modify both films. These irradiations do not have enough energy to displace atoms, but can break covalent bonds, after which the broken bond can then reform or the disrupted atoms can move so that the newly formed free radicals will be far from each other.

The free radical produced can then react with the mother macromolecule and with neighbouring macromolecules. In vacuum, the reactions result in the appearance of unsaturated structures like double bonds and aromatic rings, as well as branching and crosslinking of macromolecules. Most of the free radicals react and disappear at a high rate at room temperature but some are trapped, stabilized and remain to be detected in ESR spectra. The presence of free radicals in the exposed films shows that as the structure changes, free radicals are produced and interact with the LDPE as discussed above. The difference in the g-factors of ESR lines for samples exposed in stratosphere and on the ground shows, that the trapped free radicals are different in these films indicating that the original free radicals or/and their transformations under stratospheric conditions and ground conditions are different.

The number of unsaturated groups resulting from free radical reactions is different in the radiation damaged films exposed in the stratosphere and on the ground. The concentration of vinyl groups in the flight samples is higher than for the ground samples as expected from the generally higher level of radiation in the stratosphere than on the ground. The external surface of first film exposed in the stratosphere has more vinyl groups than the same surface of the ground sample. From the high concentration of vinyl groups in the rear surface of first film and in both surfaces of second film exposed in the stratosphere, we conclude, with the aid of TRIM simulations, that the radiation damage is caused by particles of energy greater than 10 MeV penetrating through the films. The rear surface of first film and both surfaces of second film exposed on the ground do not have an increased concentration of vinyl groups in comparison with the virgin film, from which we infer that the irradiation of these samples by high energy particles was too low to cause detectable damage.

In the presence of the atmosphere, the free radicals react with oxygen and nitrogen to give stable oxygen-containing and nitrogen-containing groups. The reactions with oxygen are well investigated and involve a number of stages. The observation of oxygen-containing groups like peroxide, hydroxyl, ketone, aldehyde, carboxyl, ether, ester, and lactone confirms that the polyethylene is oxidized [35, 37].

The XPS spectra show clear evidence of the presence of nitrogen-containing groups in polyethylene exposed in the stratosphere and on the ground. The position of the N1s line in XPS spectra corresponds to nitrogen covalently bonded to the macromolecules. The lines at 1600-1700 $cm^{-1}$ in the FTIR spectra of irradiated polyethylene can be interpreted as vibrations of $\nu$(N=O) and $\nu$(C=N) nitrogen-containing groups. However, the strong overlap of these lines with the $\nu$(C=C) and $\nu$(C=O) lines does not permit further analysis.

The activity of gaseous oxygen is much higher than nitrogen, leading to the difference in the concentrations of oxygen- and nitrogen-containing groups. The concentration of oxygen-containing groups observed in the XPS spectra of polyethylene exposed the in the stratosphere and on the ground is twice as high as the concentration of nitrogen-containing groups. This is despite the 3.5 times lower concentration of oxygen than nitrogen in the atmosphere. The difference corresponds to a 6-7 times lower activity of gaseous nitrogen than gaseous oxygen in reactions with free radicals.

The difference in atmospheric pressure explains the difference in the oxidation and nitration of the polyethylenes exposed in the stratosphere and on the ground. The lower partial pressure of oxygen in the stratosphere results in the lower concentration of oxygen-containing groups observed in the flight sample than in the sample exposed on the ground. This low concentration of oxygen-containing groups is observed together with a higher degree of radiation damage as indicated by the high concentration of vinyl groups in the stratosphere exposed sample.

The ground films contain oxygen mostly in C-O and –OH bonds, the concentration of C=O bonds being less then the concentration of C-O bonds. These products of free radical reactions correspond to the well known oxidation reactions with the sequence of products: peroxide radicals, hydroperoxide groups, oxide radicals, hydroxide groups and a further oxidation up to aldehyde and carboxyl acid groups [35, 37]:

$$RH \rightarrow R^* \rightarrow R\text{-}O\text{-}O^* \rightarrow R\text{-}O\text{-}O\text{-}H \rightarrow R\text{-}O^* \rightarrow R\text{-}OH,\ R\text{-}CHO,\ R\text{-}COOH$$

The stratosphere exposed films contain oxygen mostly in C=O bonds, the presence of C-O bonds is low and the hydroxyl groups are absent. Such a combination of oxygen-containing groups cannot be explained by known free radical reactions, where the free radical reactions are stages in the production of hydroperoxide and hydroxyl group. The significant domination of carbonyl groups can be explained, however, by a high local density of free radicals, which would favour an oxygen atom interacting with a number of unpaired electrons of a carbon atom:

$$>C^{**} + O \rightarrow >C=O$$

The low temperature and correspondingly low molecular mobility in the films irradiated in the stratosphere prevents reactions where the radicals capture hydrogen from neighbouring macromolecules. A local high density of free radicals cannot be created by low energy radiation, for example, by UV light photons, but it can be created by high energy particles, when carbon atoms are dsiplaced leaving a number of unpaired electrons on one atom. This suggests that the free radicals and the products of their reactions in polyethylene exposed in the stratosphere result from irradiation by high energy particles.

The amount of damage in polyethylene can be estimated from the known fluence of high energy ions and the measured absorbance in the FTIR spectra. From the measured vinyl and vinylene group concentrations and the known absorbance of the vinyl groups it was estimated that the polyethylene film was irradiated by $10^{13}$ ions/cm$^2$ during the 3 days in the stratosphere [38, 39]. The primary cosmic radiation cannot provide this high fluence and calculation of the fluence from satellite data (Table 1) gives $2.6x10^6$ particles/cm$^2$ during the flight, while calculations from the TIGRE counter data give $1.5x10^5$ particles/cm$^2$. This difference of fluence of $10^6$-$10^7$ is caused by the secondary particles which are generated by collisions of the primary cosmic rays with molecules in the atmosphere and the polyethylene macromolecules. Estimation of these collisions gives $10^6$-$10^9$ secondary particles per high energy proton in the primary cosmic radiation [40]. This shower of secondary particles such as electrons, muons, hadrons, and ionised atoms and molecules of the atmosphere and the polyethylene macromolecules causes the dense radiation damage in both the polyethylene films exposed in the stratosphere. If this particle shower is taken into account, satellite data give a fluence of damaging particles of $10^{12}$-$10^{15}$ particles/cm$^2$ while the TIGRE counter data gives $10^{11}$-$10^{14}$ particles/cm$^2$, both of which agree with the estimated fluence of $10^{13}$ particles/cm$^2$ calculated from the FTIR ATR spectra of the polyethylene film. This agreement shows that the structural changes in polyethylene are consistent with the expected radiation damage due to a known flux of cosmic rays. Therefore, the exposure of polymer samples in the stratosphere followed by analysis of their structural transformations can be used to estimate radiation damage effects that would occur in deep space where the effects of residual Earth atmosphere are absent or minimal. Such experiments present cost effective alternatives to space flight experiments in deep space.

Space weather satellite measurements during the period of the flight are characterised by low sun activity and low fluxes of high energy particles. Despite this low level of irradiation, the radiation damage in exposed samples of polyethylene is clearly observed. Space weather records [41] show the intensity of high energy irradiation can be up to 1000 times greater during periods of high sun activity and supernova events than during our balloon flight, and the effect of radiation damage at such times can be expected to be significantly higher.

## Conclusions

Polyethylene films were exposed in the stratosphere at 40 km altitude and on the ground as control samples during a 3 day balloon flight. Free radicals and products of their reaction, including oxidation and nitration products were found in the exposed films.

The radiation damage to the films exposed in the stratosphere was found to be caused by high energy particles of cosmic radiation and their showers. The two films, each 0.05 mm thick, exhibited damage throughout their whole thickness due to the high energy and high penetration of the cosmic rays. The oxidation and nitration of the films is consistent with the reduced atmospheric pressure and the high flux of high energy particles in the stratosphere. The film exposed on the ground had a higher degree of oxidation and nitration and less radiation damage as a consequence of the high atmospheric pressure and low flux of high energy particles at ground level. Radiation damage to the films exposed on the ground is due only to UV light.

During this flight, the flux of cosmic radiation was low corresponding to low sun and stellar activity. During periods of high solar or stellar activity the radiation damage in stratosphere is expected to be much higher.

The agreement of structural changes and expected changes estimated from cosmic ray flux data shows, that such stratospheric experiments can be used to simulate the effects of a deep space environment during interplanetary flights on polymer materials.

## Acknowledgments

The authors thank Mr. David Gregory (NASA Balloon Program) for his excellent organization of the scientific stratospheric balloon program in Australia; Mr. William Stepp, Mr. Scott Hadley and their colleagues (Columbia Scientific Balloon Facility) for an outstanding launch, the smooth flight of the balloon and the smooth landing of the payload; Prof. Allen Zych and his colleagues for useful collaboration, experimental data on cosmic radiation and friendly assistance with the payload; Prof. Ravi Sood (ADFA) for support of the mission in Alice Springs; Dr. Miriam Baltuck (CSIRO Astronomy and Space Science), Mr. Bruce Banks, Mrs. Kim de Groh and Dr. Viet Nguyen (NASA Glenn Research Center) for collaboration and support of the mission; Prof. Anne Green and Mr. Paul Harbon (University of Sydney) for financial and organizational support of the mission; Mr. Robert Davis (University of Sydney) for the preparation of the payload equipment; Dr. Elena Kosobrodova and Dr. Keith Fisher (University of Sydney) for ESR spectra, Dr. Ian Falconer (University of Sydney) for useful discussion and proof reading of the manuscript.

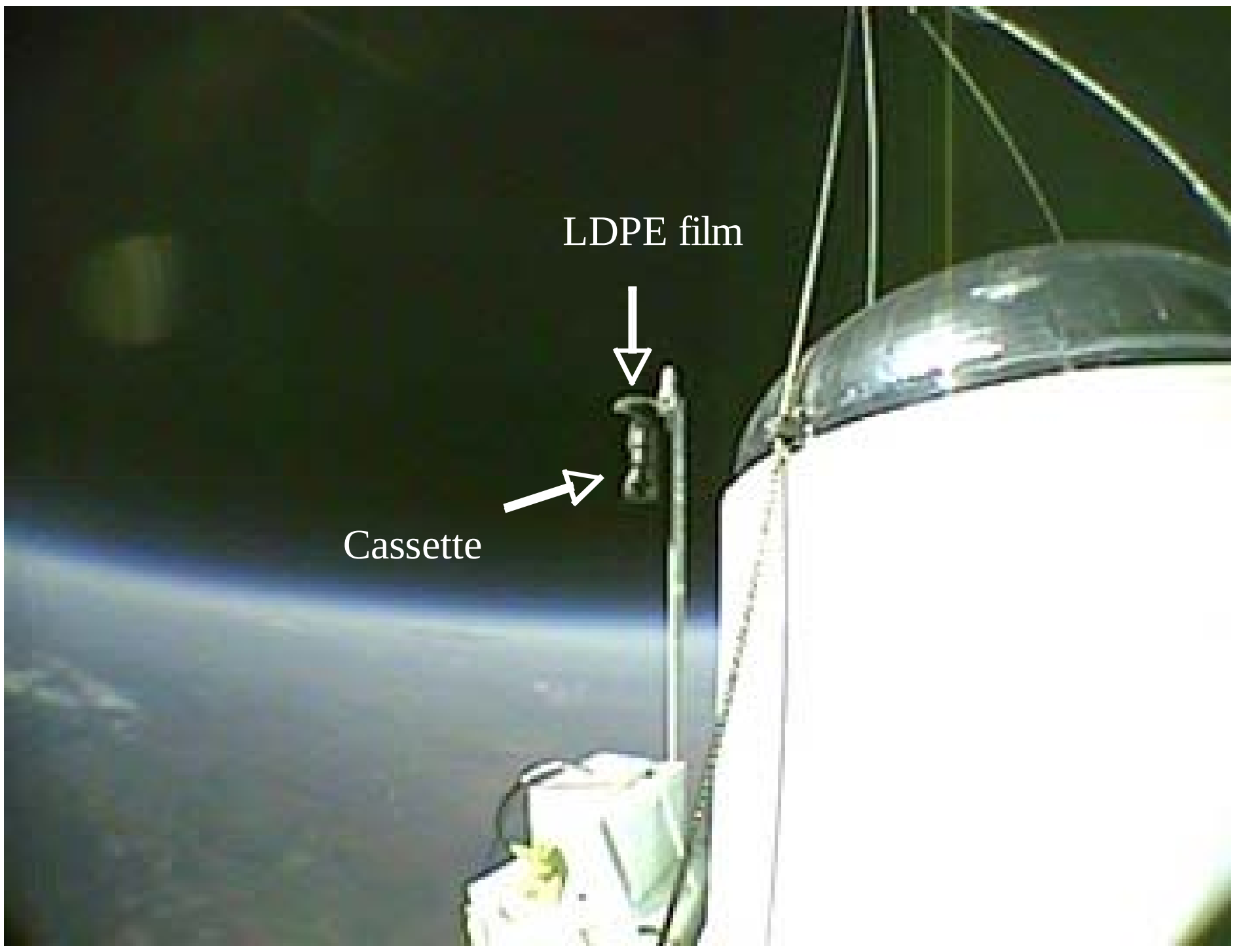

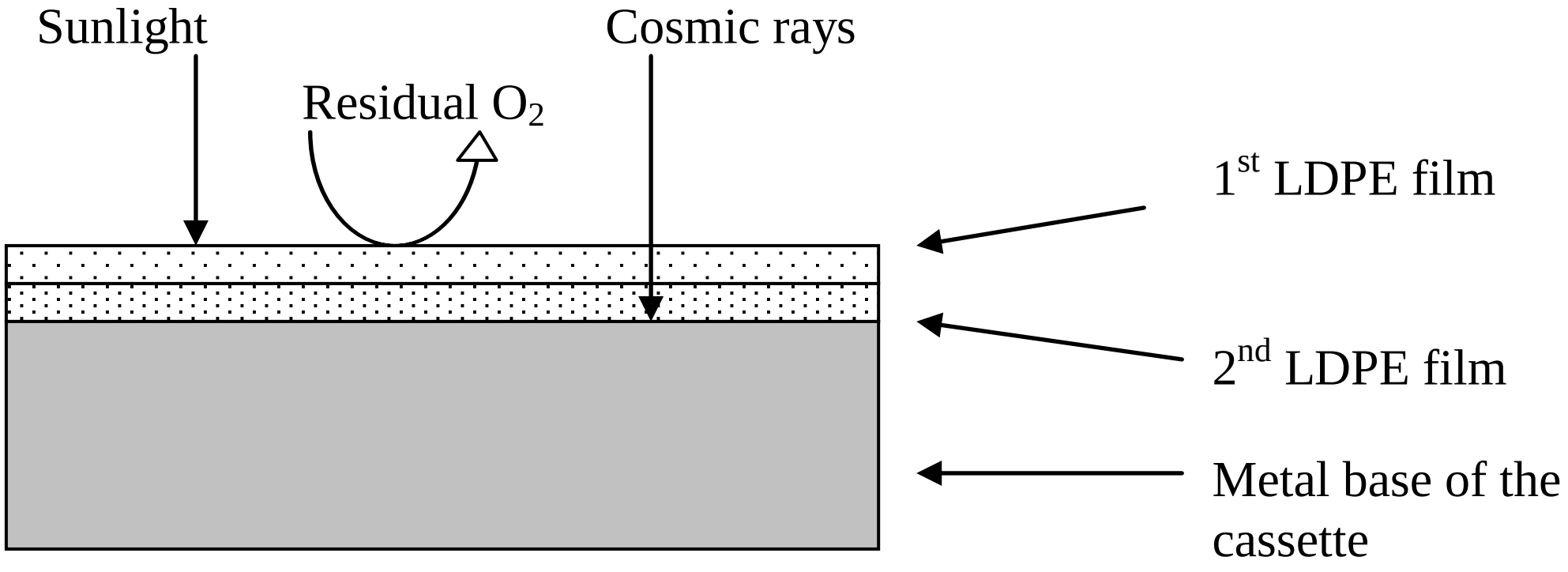

Figure 1. A telemetry screenshot showing the payload during the stratospheric flight at an altitude of 40 km. The LDPE samples are fixed on top of the cassette.

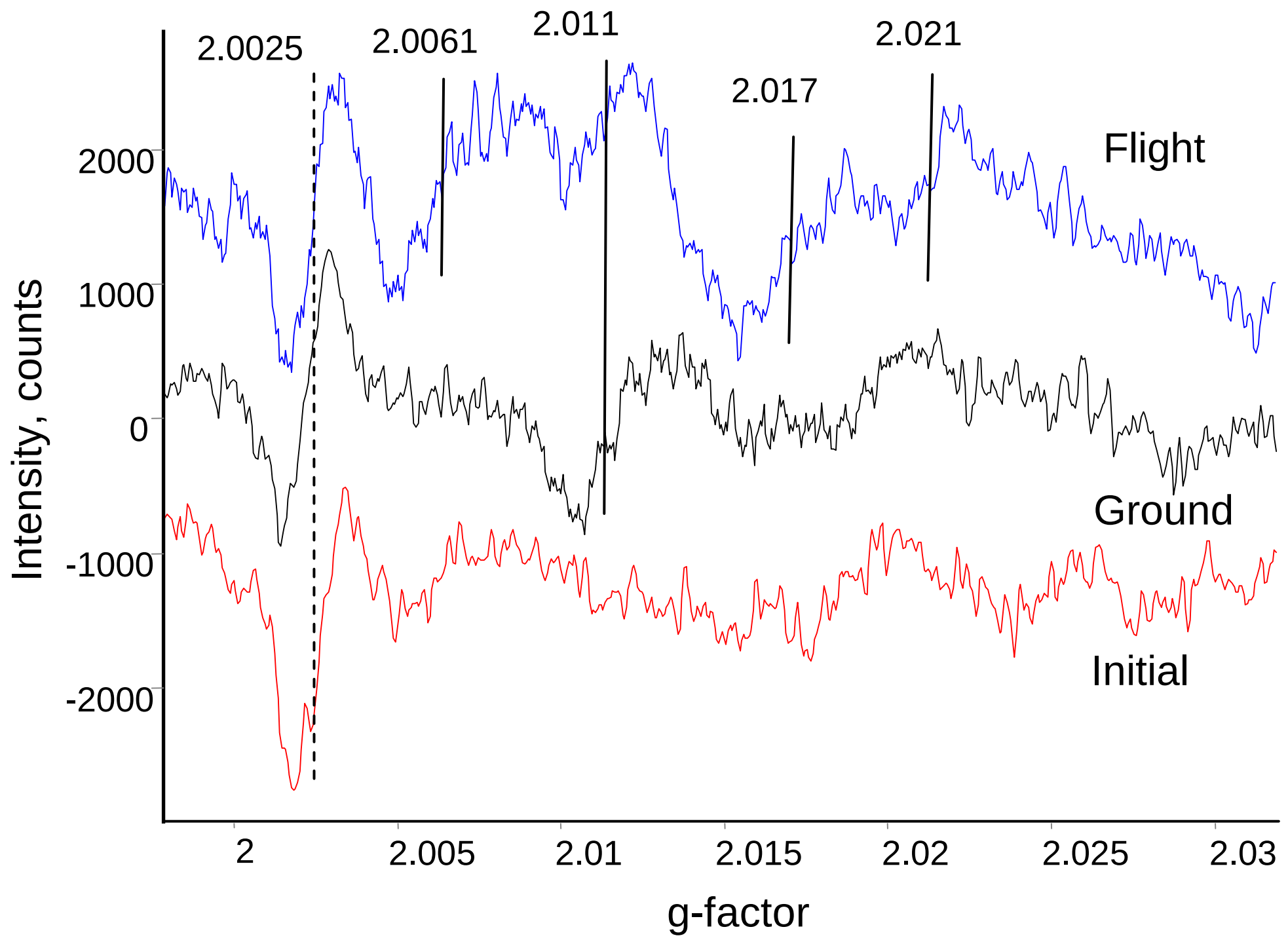

Figure 2. ESR spectra of the LDPE film after a stratospheric flight of 3 days at 40 km altitude (flight), a film exposed on the ground at the same time (ground) and an unexposed film (initial). The ESR lines with g-values of 2.0025, 2.0061, 2.011, 2.017 and 2.021 are marked.

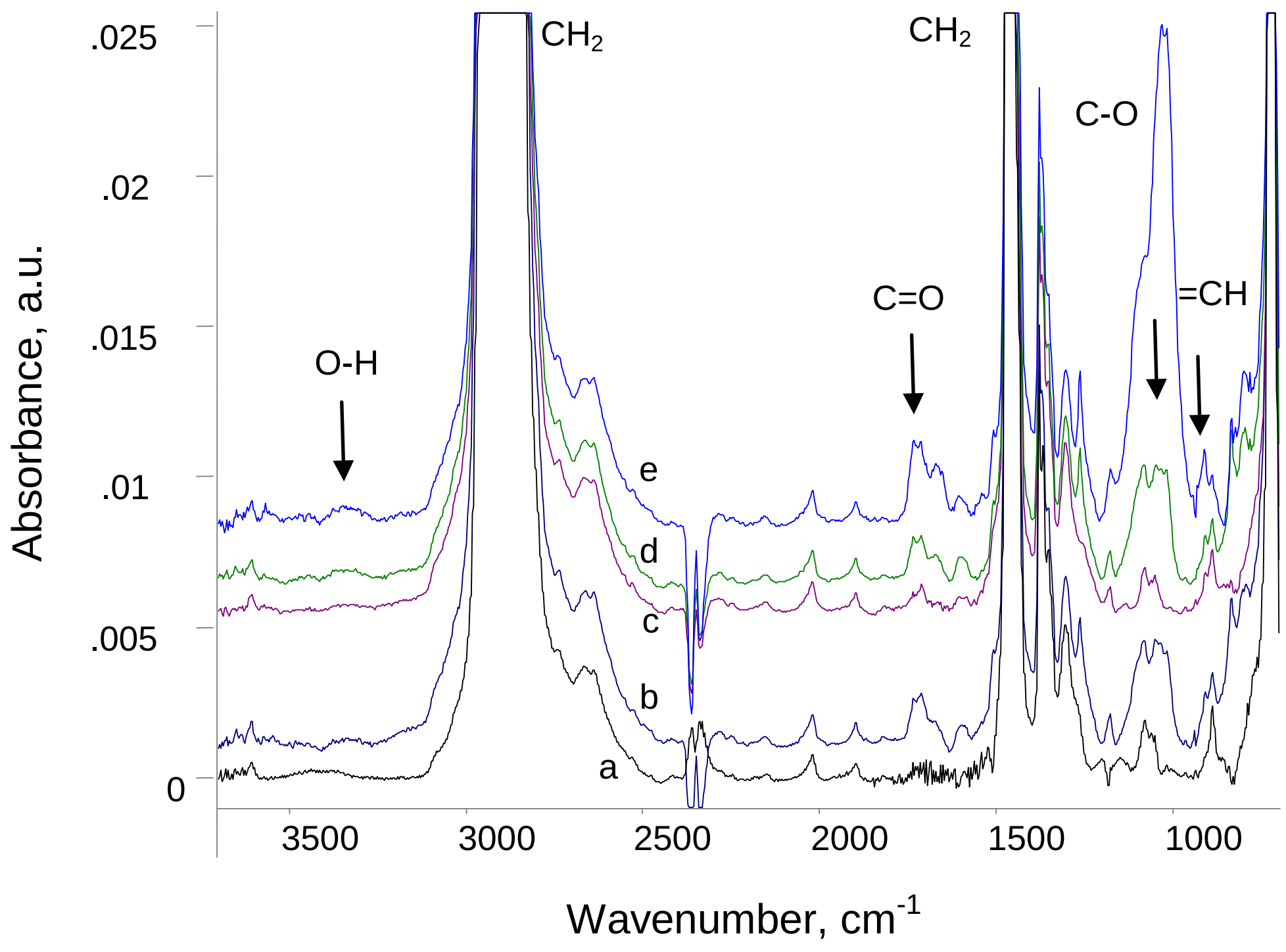

Figure 3a. FTIR ATR of LDPE after 3 days of stratospheric flight. (a) prior to the flight, (b) lower side of the second - lower - film, (c) top side of the second film, (d) lower side of the first – upper - film, (e) top side of the first film. (Flight cassette).

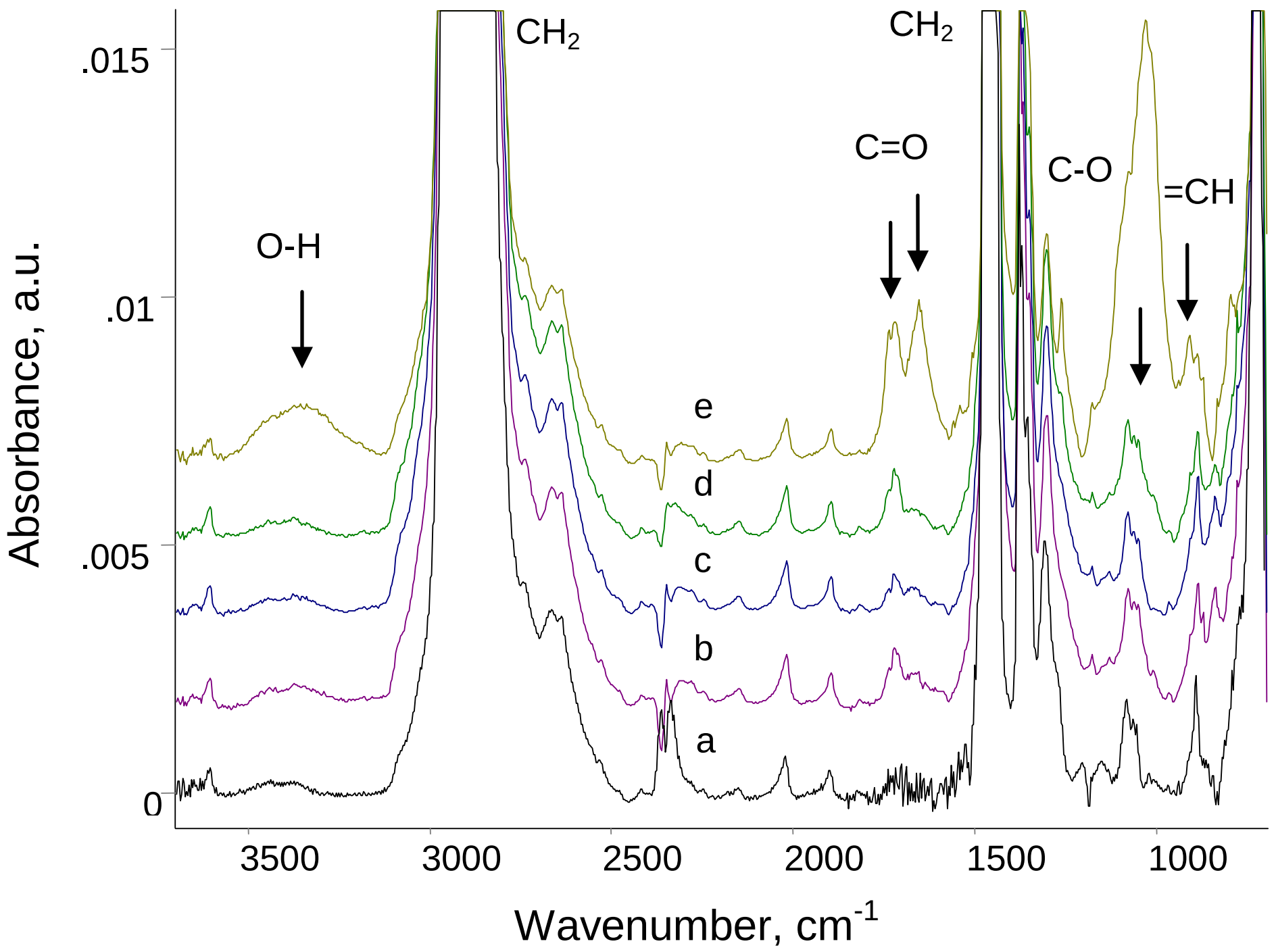

Figure 3b. FTIR ATR of LDPE after 3 days exposure on the ground. (a) prior to the exposure, (b) lower side of the second - film, (c) top side of the second film, (d) lower side of the first – upper - film, (e) top side of the first film. (Ground cassette.)

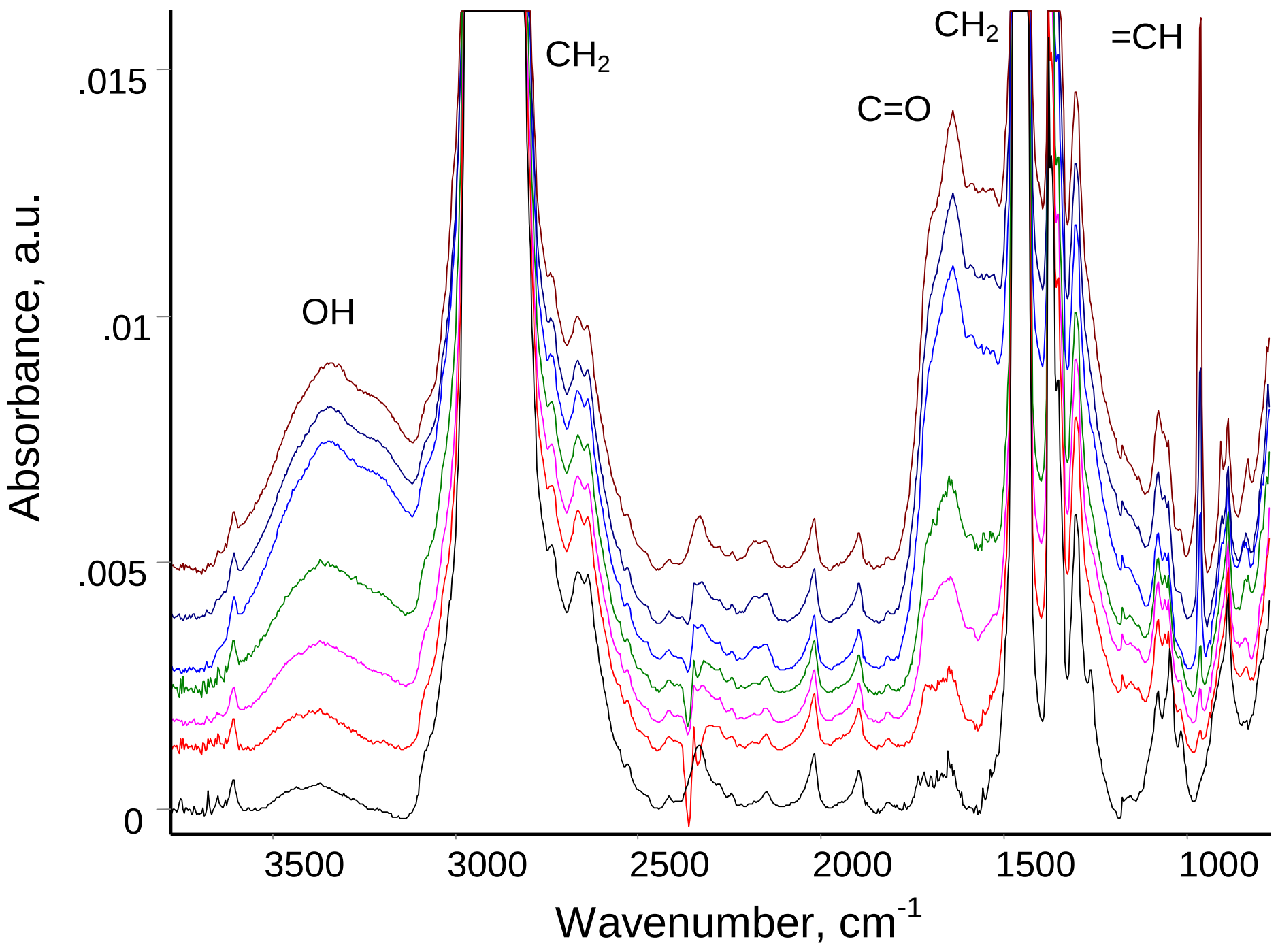

Figure 3c. FTIR ATR of LDPE after exposure in plasma for different treatment times (from bottom to top): untreated, 10 sec, 30 sec, 1 min, 5 min, 10 min, 30 min.

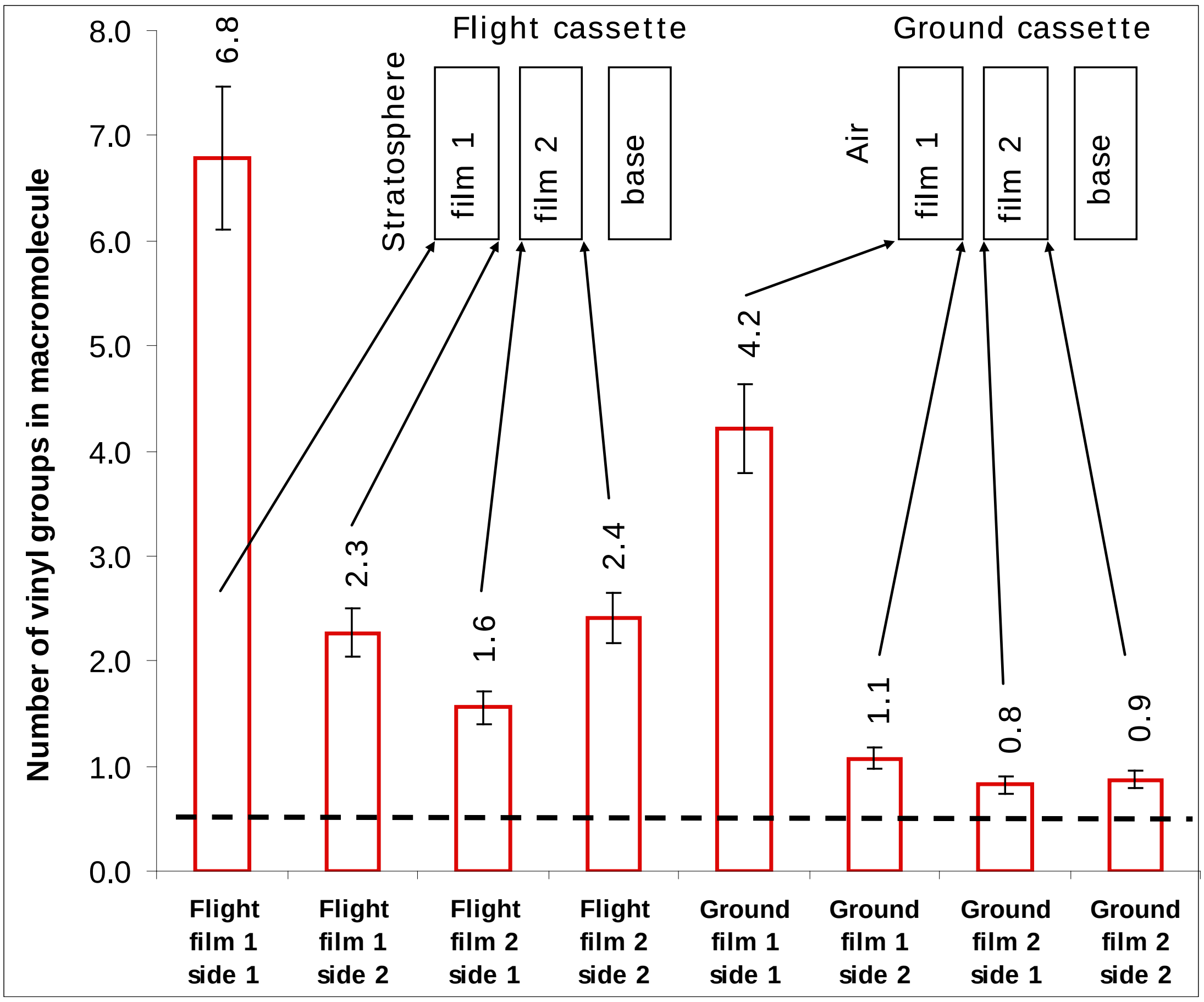

Figure 4. Relative number of vinyl groups (from FTIR ATR spectra) in LDPE exposed for 3 days in the stratosphere (flight cassette) and on the ground (ground cassette). Two films were placed on each cassette base (see Figure 1). Dashed line is for unexposed LDPE film.

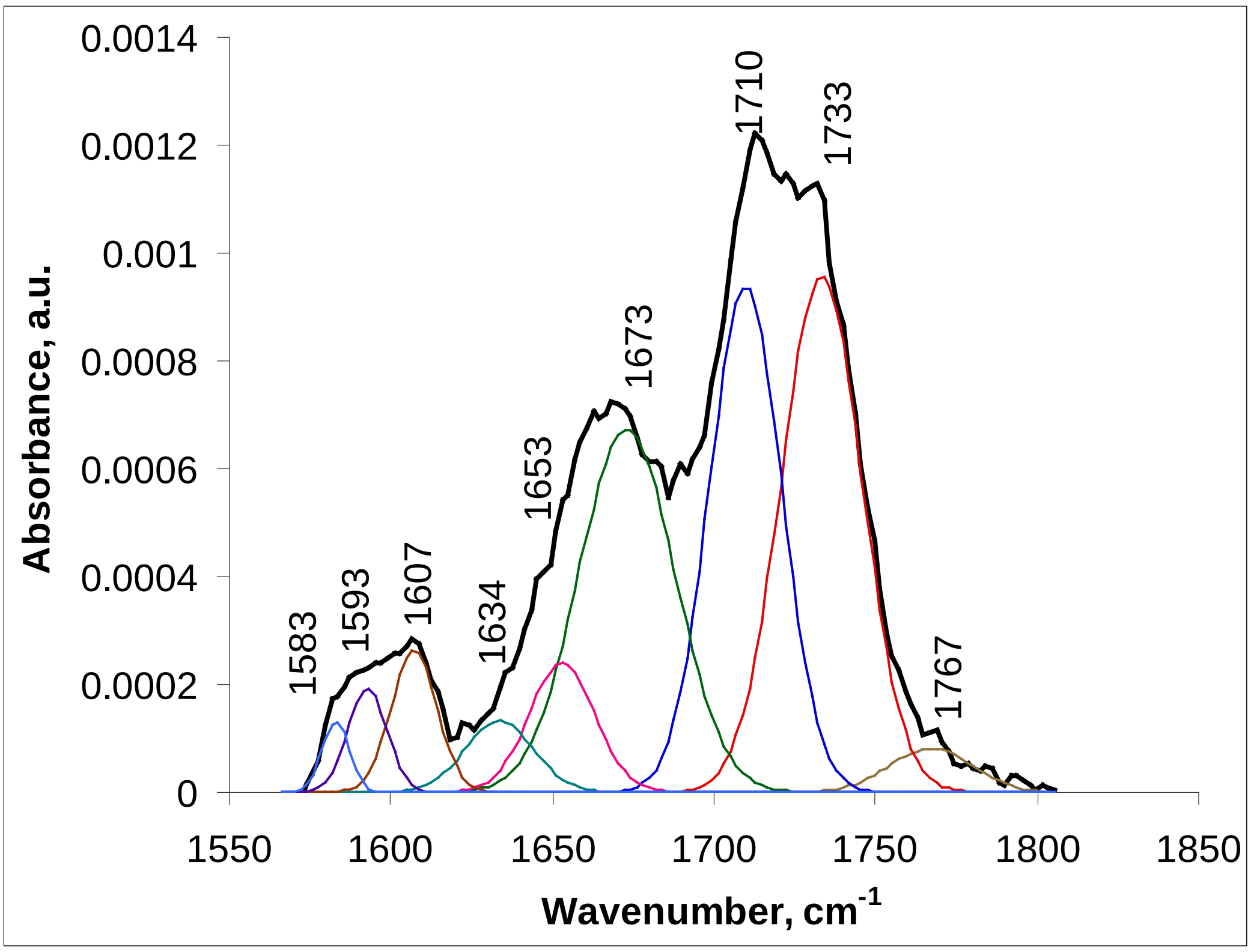

Figure 5a. Fitted FTIR ATR spectra of LDPE exposed in the stratosphere for 3 days.

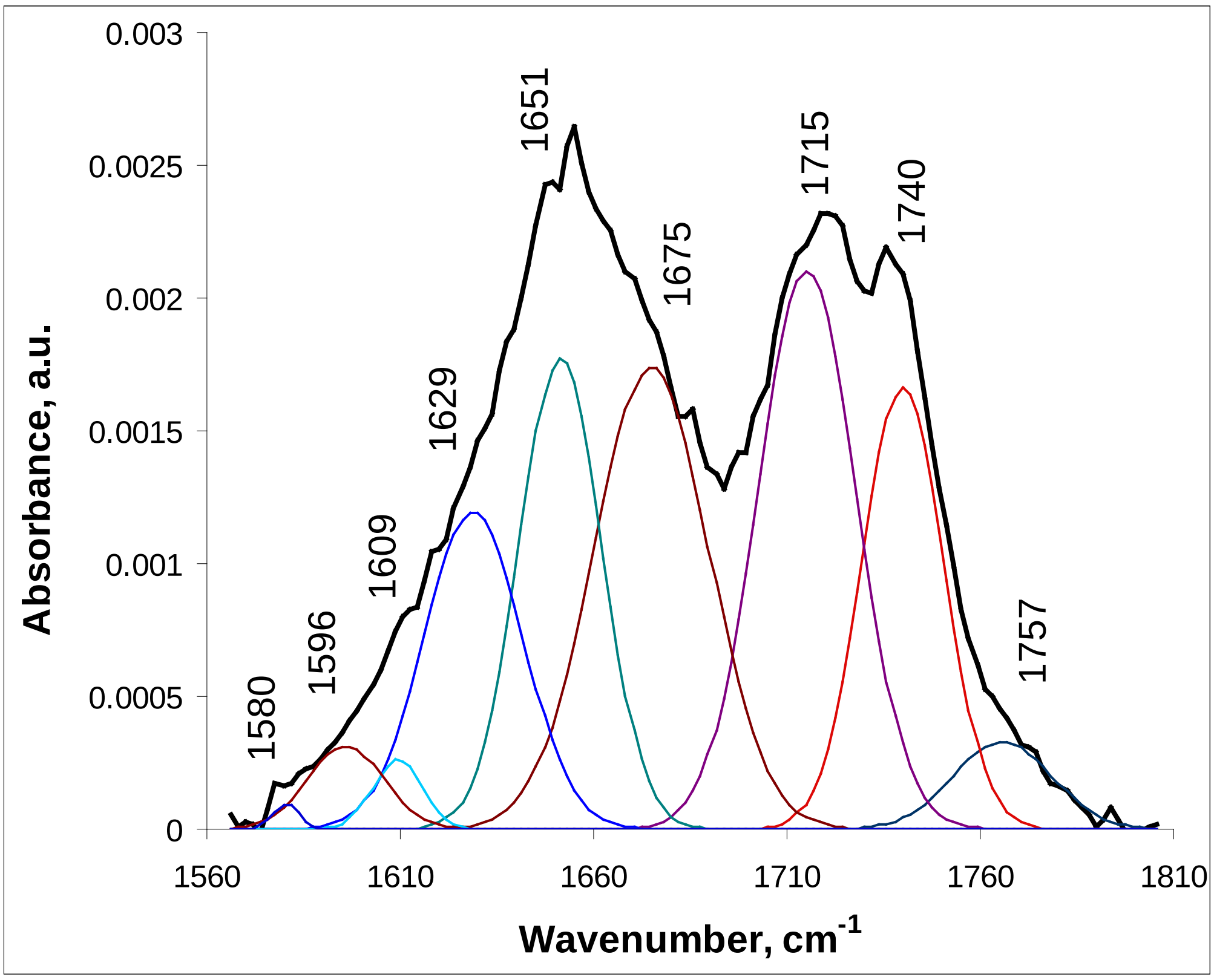

Figure 5b. Fitted FTIR ATR spectra of LDPE exposed on the ground for 3 days.

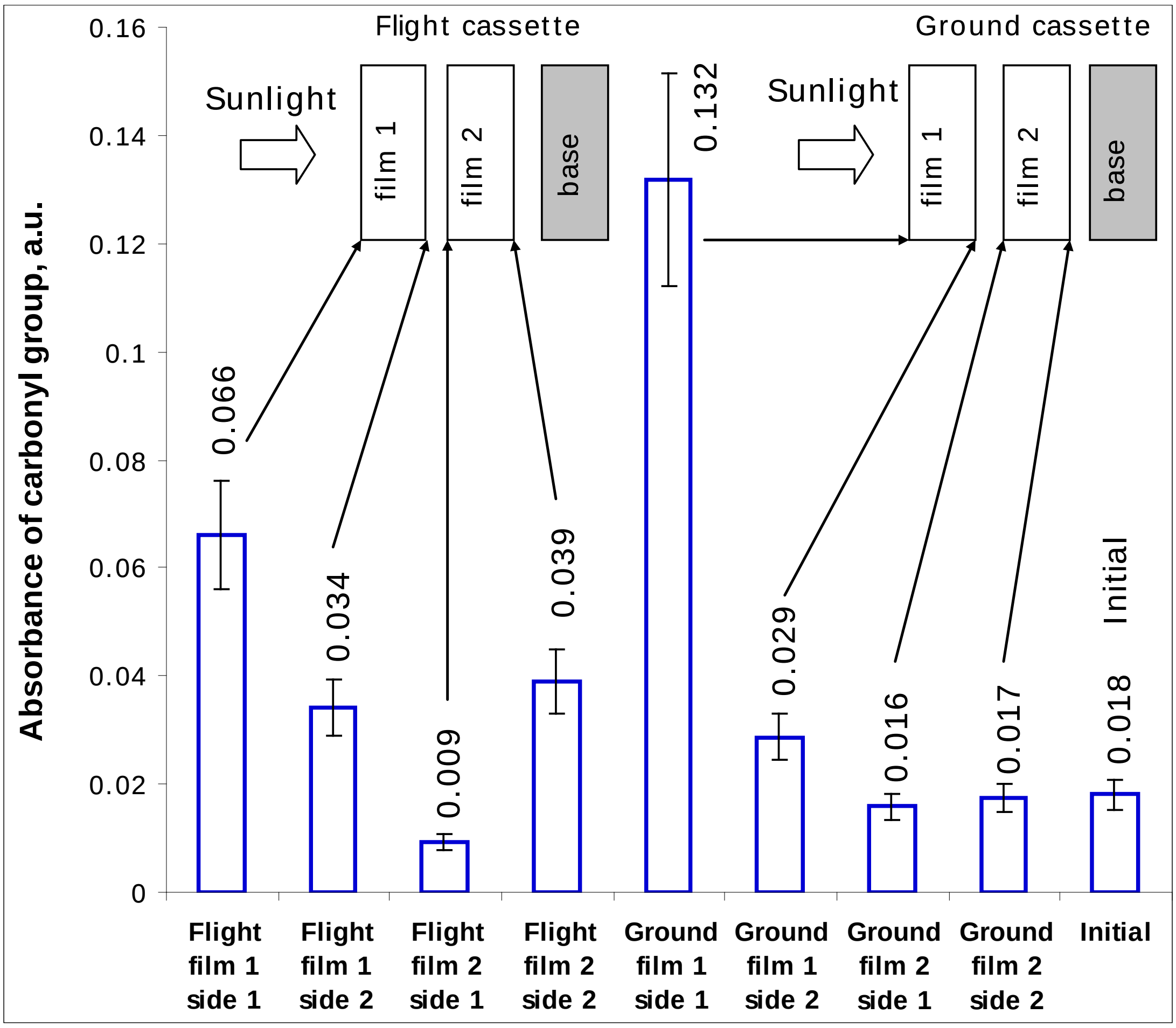

Figure 6. Integral absorbance of carbonyl group in FTIR ATR spectra of LDPE exposed for 3 days in the stratosphere (flight cassette) and on the ground (ground cassette). Two films were placed on each cassette base. Zero line is for unexposed LDPE film.

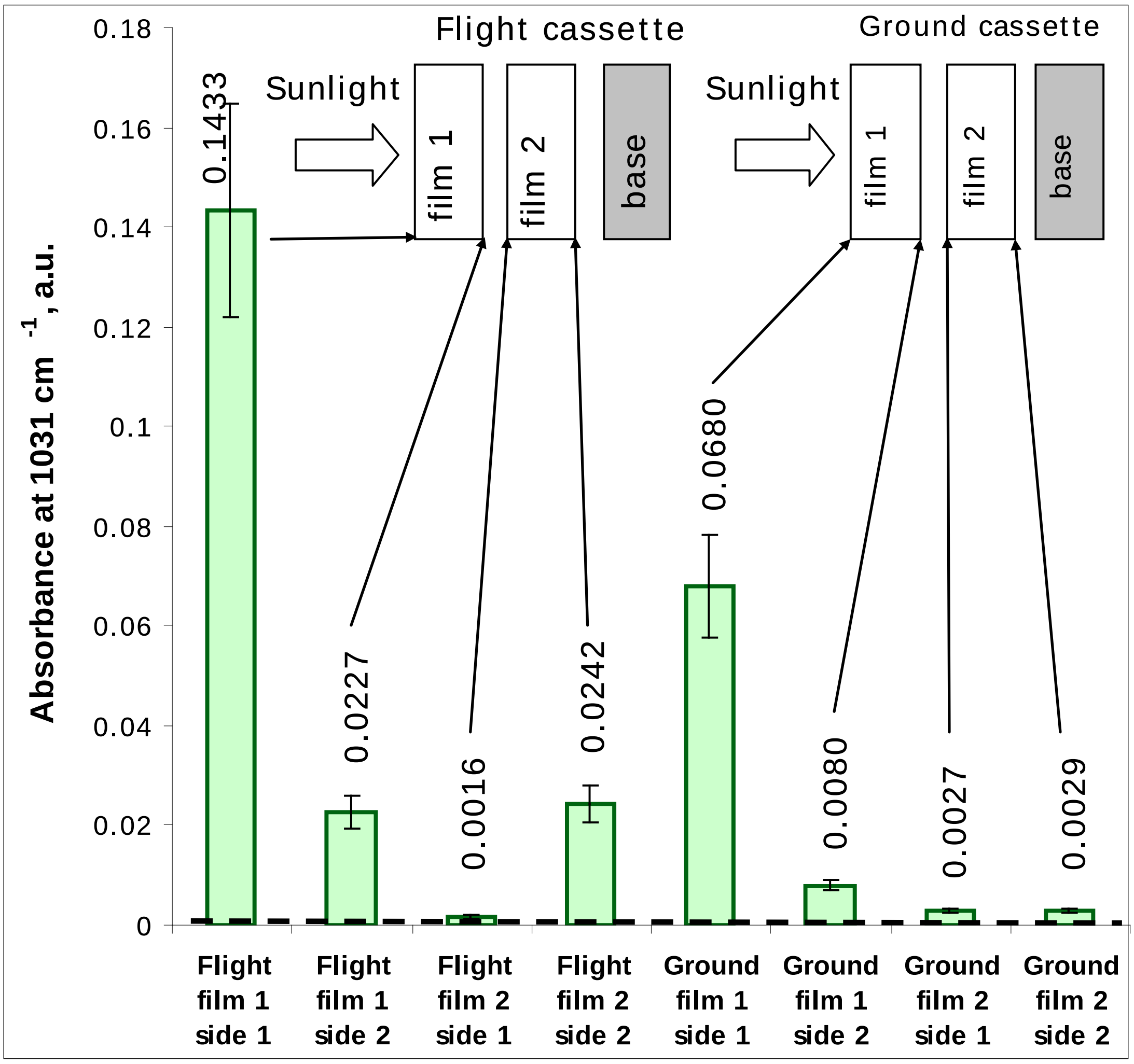

Figure 7. Absorbance of C-O group in FTIR ATR spectra of LDPE exposed for 3 days in the stratosphere (flight cassette) and on the ground (ground cassette). Two films were placed on each cassette base. Dashed line is for unexposed LDPE film.

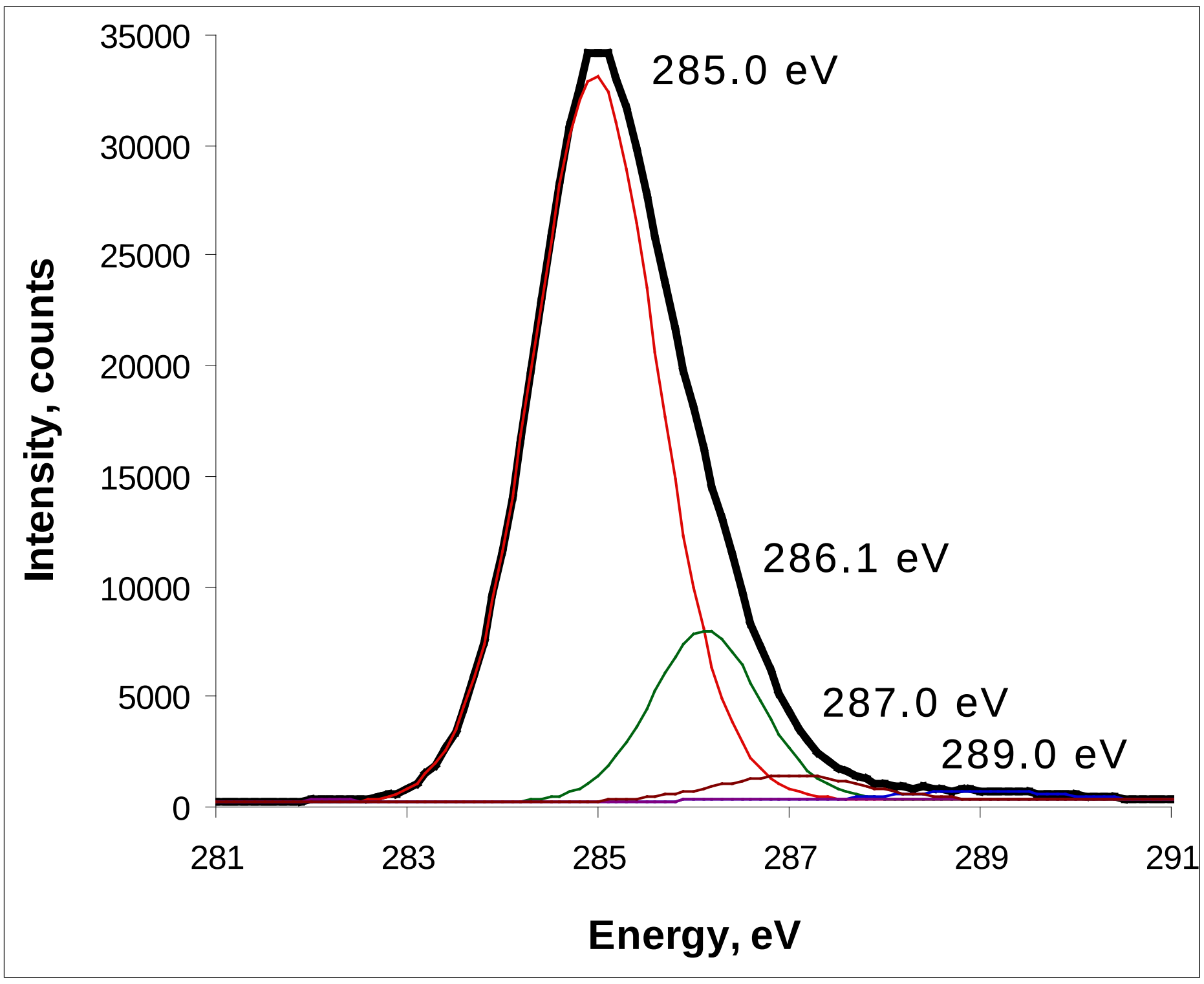

Figure 8. C1s line in XPS spectra of the external surface of LDPE film exposed for 3 days in the stratosphere.

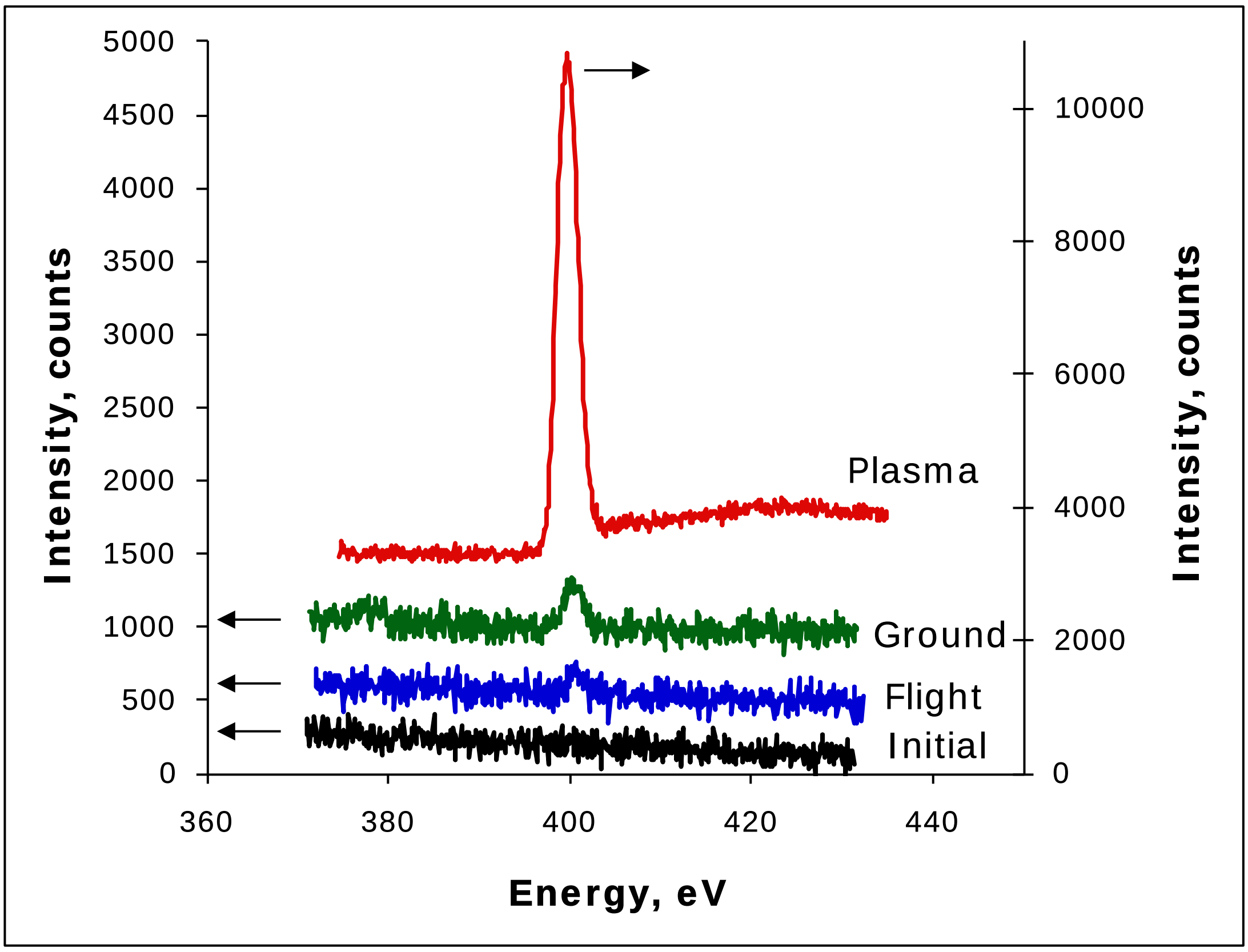

Figure 9. N1s line in XPS spectra of the external surface of LDPE film exposed for 3 days during the stratosphere flight, for 3 days on the ground, when exposed for 30 min in a nitrogen plasma, and the initial LDPE film.

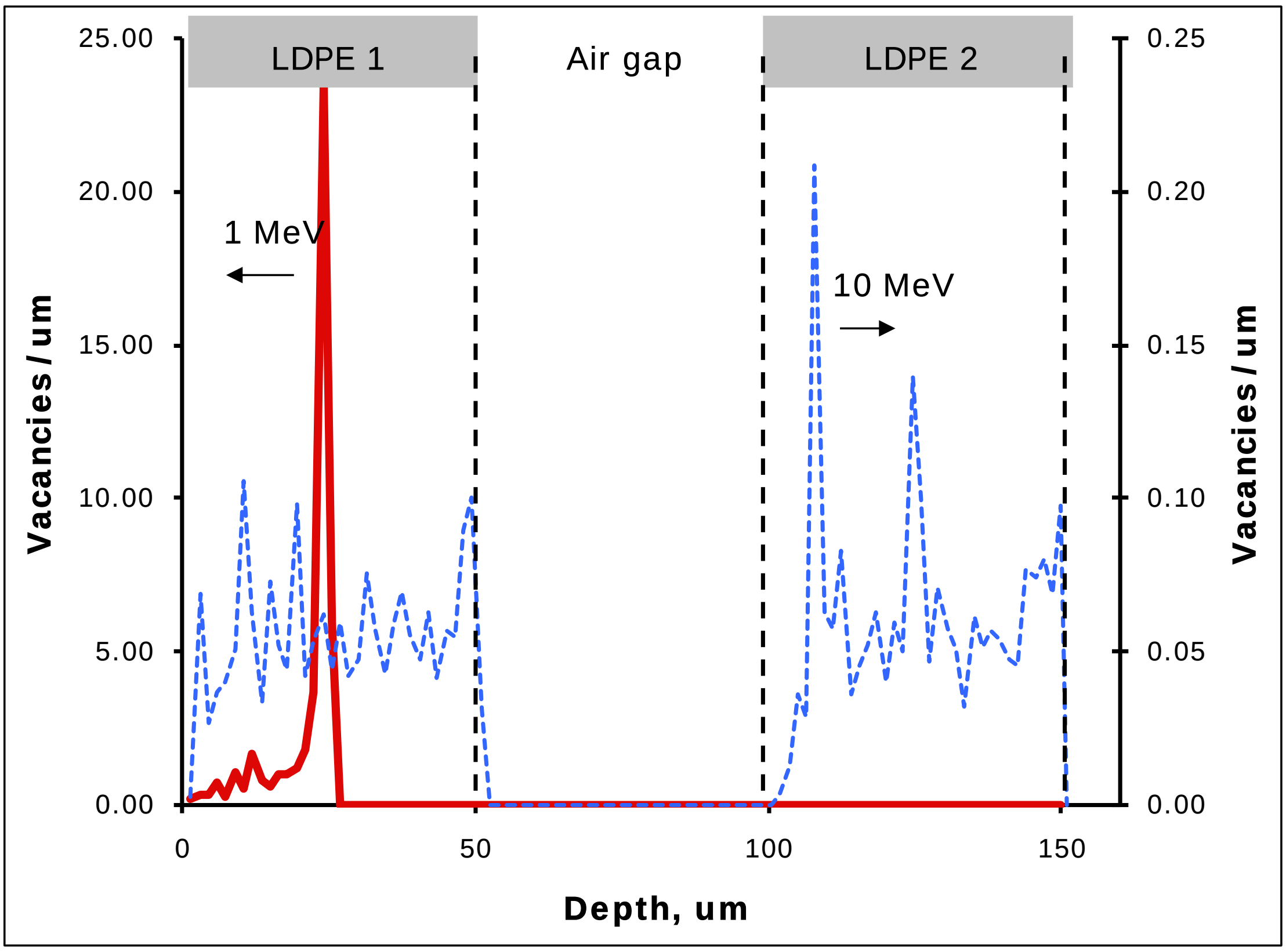

Figure 10. Dangled bonds distribution in two films of polyethylene after penetration of hydrogen ion with energies of 1 MeV (continuous line, left scale) and 10 MeV (dashed line, right scale).